\documentclass[letterpaper,aps,twocolumn,showpacs,superscriptaddress,floats,prb]{revtex4}
\usepackage{amsmath}
\usepackage{graphicx,epsfig,dsfont}
\usepackage{amssymb}
\usepackage{ulem}
\usepackage{verbatim}
\usepackage{color}
\usepackage{multirow}
\usepackage{longtable}

\newcommand{\pdag}{{\phantom{\dagger}}}

\begin{document}

\title{Weyl superconductors}

\author{Tobias Meng}
\affiliation{Institut f\"{u}r Theoretische Physik, Universit\"{a}t zu K\"{o}ln, Z\"{u}lpicher Str. 77, 50937 K\"{o}ln, Germany}
\affiliation{Institut f{\"u}r Theoretische Physik, Technische Universit{\"a}t Dresden, 01062 Dresden, Germany}

\author{Leon Balents}
\affiliation{Kavli Institute for Theoretical Physics, University of California, Santa Barbara, CA 93106, USA}

\begin{abstract}
  We study the physics of the superconducting variant of Weyl
  semimetals, which may be realized in multilayer structures
  comprising topological insulators and superconductors.  We show how
  superconductivity {can split} each Weyl node into two. The resulting
  Bogoliubov Weyl nodes can be pairwise independently controlled,
  allowing to access a set of phases characterized by different
  numbers of bulk Bogoliubov Weyl nodes and chiral Majorana surface
  modes. We analyze the physics of vortices in such systems, which
  trap zero energy Majorana modes only under certain conditions. We
  finally comment on possible experimental probes, thereby also
  exploiting the similarities between Weyl superconductors and
  2-dimensional $p+i p$ superconductors.
\end{abstract}

\date{\today}

\maketitle


\section{Introduction}

The discovery of topological insulators has stimulated a broad inquiry
into topological features of electronic energy bands.  Such features are
present not only in fully gapped systems but also in gapless ones.  Of
particular recent interest are semimetals (zero gap semiconductors) with
Fermi points, where conduction and valence bands touch at isolated
momenta in the Brillouin zone.  In three dimensions, a linear touching
between two non-degenerate bands is a Weyl point,\cite{Herring_37,volovik_book,wan_11} and is completely robust to all
perturbations which do not break translational symmetry.  Such Weyl
nodes are predicted to lead to a variety of measurable consequences,
including unusual surface states whose Fermi surface is open (``Fermi
arcs''), unusual Hall effects, and other unusual transport features, and have been studied in a number of systems.\cite{wan_11, Yang_11, Burkov_Balents_11, Hosur_11,Aji_11,Zyuzin_12,Murakami_07,Fu_11,Krempa_11,Cho_11,Young_11,Jiang_12,Delplace_12,Wang_12,Go_12}
Generically, Weyl points require a system with strong spin-orbit
coupling, and in addition the condition that the bands be non-degenerate
requires that at least either inversion or time-reversal symmetry be broken.\cite{Murakami_07,Burkov_Hook_Balents_11}
This might occur in a bulk material through magnetic order or a
non-centrosymmetric crystal structure, but Weyl points can also be
engineered.  In particular, an appropriate superlattice of alternating
normal and topological insulators has been shown to display Weyl points.\cite{Burkov_Balents_11,Halasz_Balents_12}

Weyl fermions have been discussed extensively in the context of the A
phase of ${}^3$He, which also exhibits Weyl quasiparticles.  ${}^3$He
also supports the B phase, in which quasiparticles are fully gapped, but
which nevertheless possesses interesting topological properties.\cite{volovik_book}  In
this paper, we explore the connection of Weyl semimetals to Weyl and 
topological superconductors, and more generally the effects of
superconductivity on Weyl semimetals.  Specifically we consider another
class of engineered structures, in which the normal insulator of the
aforementioned superlattice is replaced by {different layers of }a (conventional, s-wave)
superconductor.  Simple arguments show that only the Weyl semimetal
produced by time-reversal symmetry breaking, and not the one produced by
non-centrosymmetry, leads to non-trivial superconducting states.
Focusing on the former, we find that a variety of superconducting
phases, with and without Weyl points, and with varying topological
features, {can be realized depending on the degree of time-reversal symmetry breaking, and the magnitudes and phases of the superconducting order parameters in the heterostructure. We point out that the initially published version of this manuscript contained a mistake in the diagonalization of the Hamiltonian matrix, which we have corrected in the present version.\cite{thanks}}

Each of the different superconducting phases may be characterized in a number of ways, which we
discuss in the main text.  In the bulk, they may be parametrized by the
number, location, and chirality of Weyl points.  At surfaces, depending
upon both the phase and the surface orientation, one or several branches
of chiral Majorana states may be present, and extend over a varying
range of momenta in the surface Brillouin zone.  Gapless Majorana states
may also be present at vortex cores, again depending in detail upon
vortex orientation and phase.  These Majorana modes are relatives of
those proposed for use in quantum computing in two dimensional
topological superconductors.  

The paper is organized as follows. In Sec.~\ref{sec:superlattice}, we discuss how Weyl superconductors can be engineered in superlattices of superconductors and topological insulators, and derive the corresponding model. Sec.~\ref{sec:phases} is then devoted to the characterization of this Hamiltonian. We analyze how superconductivity acts on Weyl electrons in the bulk and discuss the related topological surface physics. Based thereon, a topological phase diagram is constructed. In Sec.~\ref{sec:vortices}, we analyze the physics of vortices, which under certain conditions can bind Majorana zero modes. We close with some proposals for experimental signatures based on thermal and electrical transport, which are given in Sec.~\ref{sec:exp}.


\section{Superlattice}
\label{sec:superlattice}
\begin{figure}
\centering
\includegraphics[width=0.4\columnwidth]{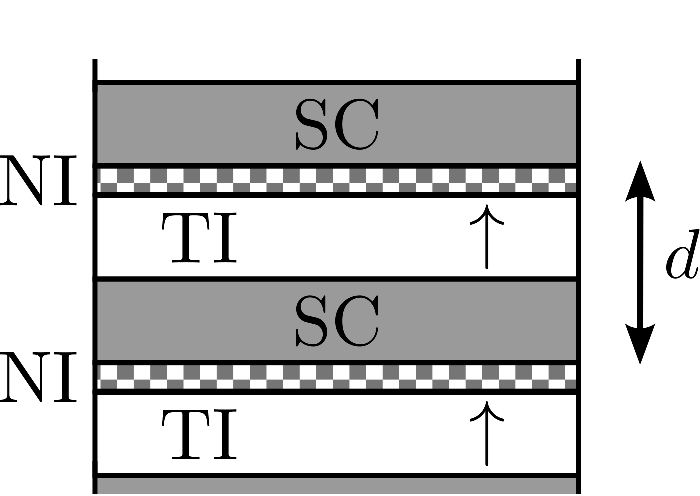}\quad\quad
\includegraphics[width=0.48\columnwidth]{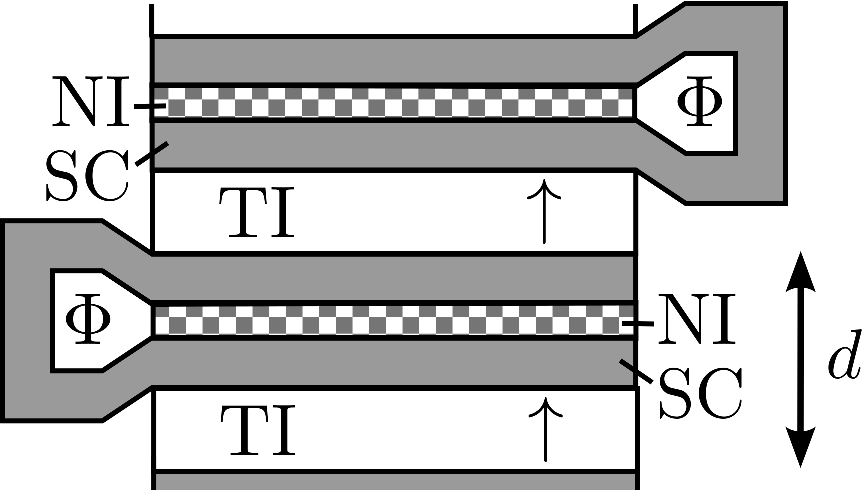}
\caption{{Heterostructures realizing different order parameter amplitudes (left) and phase differences (right) for the top and bottom surfaces of magnetically doped topological insulator layers (TI). The arrows in the TI layers depict their magnetization, which is along the superlattice axis. Different superconducting order parameter amplitudes can for example be realized by introducing normal insulating spacer layers (NI) at one of the interfaces with the superconducting layers (SC), while controllable phase differences may be achieved using Josephson junctions threaded by magnetic fluxes $\Phi$. The period of these superlattices is denoted by $d$.}}
\label{fig:heterostructure}
\end{figure}

A Weyl semimetal can be understood as an intermediate phase between
a normal insulator (NI) and a topological insulator (TI), arising due to a
perturbation of the transition between the
two.\cite{Burkov_Hook_Balents_11} One pathway to engineer Weyl
semimetals is thus the stacking of layers of topological and normal
insulators.\cite{Burkov_Balents_11} In the very same spirit, a Weyl
superconductor arises upon alternating thin layers of topological
insulators and standard s-wave BCS superconductors (SC). In such {structures, } sketched in Fig.~\ref{fig:heterostructure}, the proximity
effect induces superconductivity in the surface states of the TI
layers.

As remarked in the Introduction, to realize a Weyl semimetal requires
breaking of either time reversal or inversion symmetry.  In the bulk of
this paper we focus on the time reversal symmetry breaking case, as it
leads to much more non-trivial results in the presence of
superconductivity.  Indeed, in Appendix \ref{append:gap_or_no_gap}, we show that when
inversion symmetry is instead broken, while time reversal is preserved,
{the} superconducting proximity effect leads directly to a gapped, trivial
phase.  Specifically we model the topological insulator layers by a
single Dirac node per surface, with an imposed Zeeman splitting which may be
considered to arise from an exchange coupling to randomly distributed magnetic impurities ferromagnetically polarized
perpendicular to the TI layers, while the coupling between the TI surface modes and the magnetic field of the impurities is neglected as usual in such systems.\cite{Liu_09, Yu_10, Pesin_11, Rosenberg_12, Sau_10}  {The superconducting layers induce a pairing amplitude for the TI surface states by proximity. The strength and phase of this proximity-induced superconductivity on the top and bottom surfaces of the TI layers are in general tunable by additional spacing layers, and magnetic fluxes, as shown in Fig.~\ref{fig:heterostructure}}.  Furthermore,
neighboring {TI} surface layers are tunnel coupled. Longer range tunneling is
assumed to be negligible.  

\subsection{Model Hamiltonian}
\label{sec:model-hamiltonian}

Working in units of $\hbar = 1$, we model the system by the Hamiltonian
\begin{align}
\label{eq:heterostructure_ham1}
H=& \sum_{\vec{k}_\perp,i,j} c_{\vec{k}_\perp i}^\dagger \, \mathcal{H}_{ij} \, c_{\vec{k}_\perp j}^\pdag + H_{SC}\text{ ,}\\
\mathcal{H}_{ij} =& \,v_F \,\tau^z\,\left(\hat{z}\times\vec{\sigma}\right)\cdot \vec{k}_\perp \,\delta_{i,j} + m \,\sigma^z \,\delta_{i,j}+ t_S \,\tau^x \,\delta_{i,j}\nonumber \\
& + \frac{1}{2} \,t_D \,\tau^+ \,\delta_{i,j+1} + \frac{1}{2} \,t_D \,\tau^-\, \delta_{i,j-1}\label{eq:heterostructure_hamij1}\\
H_{SC} =&\sum_{\vec{k}_\perp,i}  \left({\Delta^{\text{top}}} c_{\vec{k}_\perp \uparrow i}^{\text{top}}{}^\dagger c_{-\vec{k}_\perp \downarrow i}^{\text{top}}{}^\dagger + {\Delta^{\text{bot.}}}c_{\vec{k}_\perp \uparrow i}^{\text{bot.}}{}^\dagger c_{-\vec{k}_\perp \downarrow i}^{\text{bot.}}{}^\dagger \right)\nonumber\\
&+ \text{h.c.} \text{ ,} \label{eq:heterostructure_sc1}
\end{align}
where $c_{\vec{k}_\perp i} = (c_{\vec{k}_\perp \uparrow i}^{\text{top}},
c_{\vec{k}_\perp \downarrow i}^{\text{top}}, c_{\vec{k}_\perp \uparrow
  i}^{\text{bot.}}, c_{\vec{k}_\perp \downarrow i}^{\text{bot.}} )^T$
comprises annihilation operators for electrons of spin up and down in
the top and bottom surfaces of layer $i$ with in-plane momentum
$\vec{k}_\perp$. The unit vector along the perpendicular axis is
$\hat{z}$. The Fermi velocity of the Dirac nodes is $v_F$, for
simplicity considered to be the same on each surface, and Pauli matrices
$\vec{\sigma}$ act on the real spin. The additional pseudo spin for the
top/bottom surface degree of freedom denoted by the Pauli matrices
$\vec{\tau}$. The Zeeman mass (half the Zeeman splitting) is $m$, the
tunneling between top and bottom surface of the same TI layer is denoted
by $t_S$, and the tunneling between different TI layers is $t_D$. The
proximity induced superconductivity is characterized by {$\Delta^{\text{top}} =
(|\Delta|+\delta\Delta)e^{i\varphi/2}$ and $\Delta^{\text{bot.}} =
(|\Delta|-\delta\Delta)e^{-i\varphi/2}$, with $|\Delta|>0$ denoting the average pairing amplitude, $\delta\Delta$ characterizing the pairing amplitude difference (with $|\delta\Delta|<|\Delta|$), and $\varphi$ being the
superconducting phase difference between top and bottom layers (note that a global superconducting phase can be absorbed into the operators). A finite $\delta\Delta$ can be realized by controlling the interface roughnesses, or may arise due to additional normal insulating spacer layers between one of the surface of the topological insulator layers and the superconductors, see Fig.~\ref{fig:heterostructure} (left). Finite phase differences between the top and bottom layers can be created by Josephson junctions in which the phase difference is controlled by magnetic fields, see Fig.~\ref{fig:heterostructure} (right). As we will show now, \textit{either} a finite phase difference, \textit{or} a different pairing amplitude on top and bottom surfaces is already enough to split the Weyl nodes into Bogoliubov Weyl nodes.}

{\subsection{Normal state}}
\label{sec:normal-state}

We {first analyze the normal state Hamiltonian with $|\Delta|=0=\delta\Delta$, and }proceed by Fourier transforming the Hamiltonian along $\hat{z}$, where the superlattice constant is $d$. After a canonical transformation
\begin{equation}
\sigma^\pm \to \tau^z \sigma^\pm\, , \quad\quad \tau^\pm \to \sigma^z \tau^\pm\, ,
\end{equation}
and subsequent diagonalization in the $\vec{\tau}$ subspace, the Hamiltonian reads
\begin{align}
\label{eq:heterostructure_ham2}
H=& \sum_{\vec{k},l=\pm}\tilde{c}_{\vec{k}l}^\dagger \, \mathcal{H}_l \, \tilde{c}_{\vec{k}l}^\pdag \text{ ,}\\
\mathcal{H}_\pm =& \,v_F \,\left(\hat{z}\times\vec{\sigma}\right)\cdot \vec{k} + M_\pm(k_z) \,\sigma^z \text{ ,} \label{eq:heterostructure_ij2}\\
M_\pm(k_z) =& m \pm \sqrt{t_S^2 + t_D^2 + 2 \, t_S t_D \cos\left(k_z d\right)}\text{ ,}
\end{align}
where  $\tilde{c}_{\vec{k} \pm} = (\tilde{c}_{\vec{k} \uparrow \pm}^\pdag, \tilde{c}_{\vec{k} \downarrow \pm}^\pdag)^T$ is now composed of the appropriate eigenoperators resulting from the diagonalization in the $\vec{\tau}$-subspace, and $\vec{k}$ being the 3-dimensional momentum.

A Weyl node corresponds to the vanishing of an eigenenergy of
\eqref{eq:heterostructure_ham2} for one momentum. In the case without
superconductivity analyzed in Ref. \onlinecite{Burkov_Balents_11}, and
assuming without loss of generality that $m>0$ and $t_S/t_D > 0$, two
Weyl nodes of opposite chirality can appear in the spectrum of
$\mathcal{H}_-$. They are located at $\vec{k} = \left(0,0,\pi/d\pm
  k_0\right)^T$ with
\begin{equation}
k_0 = \frac{1}{d}\arccos\left(1-\frac{m^2-(t_S-t_D)^2}{2 \, t_S \,
    t_D}\right) ,
\end{equation}
as long as the condition
\begin{equation}
m_{c1}^2 = \left(t_S-t_D\right)^2 < m^2 < \left(t_S+t_D\right)^2 = m_{c2}^2 \label{eq:mc_i}
\end{equation}
is fulfilled. For $m^2 < \left(t_S-t_D\right)^2$, $\mathcal{H}_-$ describes a trivial insulator, while $m^2> \left(t_S+t_D\right)^2$ corresponds to a quantum anomalous Hall insulator. The Hamiltonian $\mathcal{H}_+$, on the other hand, always describes a trivial insulator. It is adiabatically connected to the case $m = 0$ that is topologically trivial.

\subsection{Superconducting state}
\label{sec:superc-state}

For the superconducting case, we keep $m>0$ and $t_S/t_D > 0$, although
a different choice does not change our results qualitatively. Technically,
superconductivity is taken into account by introducing a particle-hole
pseudospin on which the Pauli matrices $\vec{\kappa}$ act, as well as
the corresponding Nambu spinors.  {Using 

\begin{align}
{\Psi}_{\vec{k}}^T = (&c_{\vec{k},\uparrow}^{\rm top},c_{\vec{k},\downarrow}^{\rm top},e^{i k_zd/2}\,{c}_{\vec{k},\uparrow}^{\rm bot.},e^{i k_zd/2}\,{c}_{\vec{k},\downarrow}^{\rm bot.},\nonumber\\
&c_{\vec{-k},\downarrow}^{\rm top}{}^\dagger,c_{\vec{-k},\uparrow}^{\rm top}{}^\dagger,e^{i k_zd/2}\,{c}_{\vec{-k},\downarrow}^{\rm bot.}{}^\dagger,e^{i k_zd/2}\,{c}_{\vec{-k},\uparrow}^{\rm bot.}{}^\dagger)
\end{align}

we can write the Hamiltonian as
\begin{align}
H & = \frac{1}{2}\sum_{\vec{k}} \Psi_{\vec{k}}^\dagger \,{\mathcal{H}}({\vec{k}}) \,\Psi_{\vec{k}}^\pdag ,\\
\end{align}
with
\begin{align}
\mathcal{H}({\vec{k}})&=v_F(\sigma_x k_y - \sigma_y k_x )\,\tau_z\,\mathds{1}_\kappa+m\,\sigma_z\,\mathds{1}_\tau\,\mathds{1}_\kappa \nonumber\\
&+ (t_s+t_d)\cos\left(\frac{k_z d}{2}\right)\,\mathds{1}_\sigma\,\tau_x\,\kappa_z\nonumber\\
&+(t_s-t_d)\sin\left(\frac{k_z d}{2}\right)\,\mathds{1}_\sigma\,\tau_y\,\kappa_z\nonumber\\
&+|\Delta|\cos(\varphi/2)\,\sigma_z\,\mathds{1}_\tau\,\kappa_x+\delta\Delta\cos(\varphi/2)\,\sigma_z\,\tau_z\,\kappa_x\nonumber\\
&-|\Delta|\sin(\varphi/2)\,\sigma_z\,\tau_z\,\kappa_y-\delta\Delta\sin(\varphi/2)\sigma_z\mathds{1}_\tau\kappa_y.\label{eq:initial}
\end{align}
This Hamiltonian can be diagonalized with the canonical transformations
\begin{align}
\sigma^\pm \to \tau^z \sigma^\pm\, , \quad\quad \tau^\pm \to \sigma^z \tau^\pm\, ,
\end{align}
followed by
\begin{align}
\tau^\pm \to \kappa^z \tau^\pm\, , \quad\quad \kappa^\pm \to \tau^z \kappa^\pm\, ,
\end{align}
which yields
\begin{align}
\mathcal{H}({\vec{k}})&=v_F(\sigma_x k_y - \sigma_y k_x )\,\mathds{1}_\tau\,\mathds{1}_\kappa+m\,\sigma_z\,\mathds{1}_\tau\,\mathds{1}_\kappa\nonumber\\
&+ (t_s+t_d)\cos\left(\frac{k_z d}{2}\right)\,\sigma_z\,\tau_x\,\mathds{1}_\kappa \nonumber\\
&+(t_s-t_d)\sin\left(\frac{k_z d}{2}\right)\,\sigma_z\,\tau_y\,\mathds{1}_\kappa \nonumber\\
&+\sigma_z\Bigl(|\Delta|\cos(\varphi/2)\,\tau_z\,\kappa_x+\delta\Delta\cos(\varphi/2)\,\mathds{1}_\tau\,\kappa_x\nonumber\\
&-|\Delta|\sin(\varphi/2)\,\mathds{1}_\tau\,\kappa_y-\delta\Delta\sin(\varphi/2)\,\tau_z\,\kappa_y\Bigr).
\end{align}
For the most simple case $\delta\Delta=0=\varphi$, this Hamiltonian has a two-fold degenerate spectrum. For general finite values of $\delta\Delta$ and/or $\varphi$, corresponding to a generic heterostructure without perfect symmetry between the top and bottom surfaces of the topological insulator layers, the spectrum is non-degenerate, and each Weyl node is split into a pair of Bogoliubov Weyl nodes. Therefore, a multilayer structure of topological insulators and BCS $s$-wave superconductors will generally exhibit split Bogoliubov Weyl nodes. {Similar results have also been obtained for intrinsic superconductivity in Weyl semimetals, see for instance Refs.~\onlinecite{gong_11,sau_12,cho_12,bednik_15}, which also demonstrate the existence of Bogoliubov Weyl nodes in systems that might experimentally be more accessible than topological insulator heterostructures.} To simplify the subsequent discussion of the physical implications of Weyl nodes splitting into pairs of Bogoliubov Weyl nodes, we from now on focus on the case of a heterostructure with a $\pi$-phase difference between the top and bottom surfaces of each TI layer, $\delta\Delta =0$ and $\varphi=\pi$, for which the algebra is simplest. Because the $\vec{\tau}$ and $\vec{\kappa}$ subspaces are then decoupled, we can diagonalize the $\vec{\tau}$-subspace first.}
Using $\psi_{\vec{k}} = (c_{\vec{k}
  \uparrow -}^\pdag, c_{\vec{k} \downarrow -}^\pdag, c_{-\vec{k}
  \downarrow -}^\dagger, c_{-\vec{k} \uparrow -}^\dagger)^T$, the
$\tau^z = -1$ sector of the Hamiltonian can be recast into the form
\begin{align}
H_- = \frac{1}{2}\sum_{\vec{k}} \psi_{\vec{k}}^\dagger &\biggl[\left(v_F \,\left(\hat{z}\times\vec{\sigma}\right)\cdot \vec{k} + M_-(k_z)\sigma^z\right) \mathds{1}_{\vec{\kappa}}\nonumber\\
&-|\Delta|\,\sigma^z\,\kappa_y\biggr]\psi_{\vec{k}}^\pdag \text{ .}\label{eq:h_-}
\end{align}
Diagonalization of the $\vec{\kappa}$ subspace yields
\begin{equation}
H_- = \frac{1}{2} \sum_{\vec{k}, n=\pm} \Phi_{\vec{k},n}^\dagger \,\mathcal{H}_-^{n \Delta}\, \Phi_{\vec{k},n}^\pdag \text{ ,}\label{eq:ham_-}
\end{equation}
where
\begin{align}
\mathcal{H}^{\pm\Delta}_- =&\,v_F \,\left(\hat{z}\times\vec{\sigma}\right)\cdot \vec{k} + M^{\pm\Delta}_-(k_z) \,\sigma^z \label{eq:bogolibov_ham_pm}\text{ ,}\\
M^{\pm\Delta}_-(k_z) =& \left(m \pm |\Delta|\right) - \sqrt{t_S^2 + t_D^2 + 2 \, t_S t_D \cos\left(k_z d\right)}\text{ ,} \label{eq:m_pm_delta}
\end{align}
and 
\begin{subequations}
\begin{align}
\Phi_{\vec{k},+} &= \left(d_{\vec{k}}^\pdag,d_{-\vec{k}}^\dagger\right)^T\text{ ,} \quad\quad \Phi_{\vec{k},-} = \left(f_{\vec{k}}^\pdag,f_{-\vec{k}}^\dagger\right)^T \text{ ,}\\
d_{\vec{k}} &= \frac{1}{\sqrt{2}} \left(e^{-i \pi/4} \, c_{\vec{k} \uparrow-}^\pdag + e^{+i \pi/4} \, c_{-\vec{k} \downarrow-}^\dagger\right) \text{ ,} \label{eq:dk} \\
f_{\vec{k}}&= \frac{1}{\sqrt{2}i} \left(e^{-i \pi/4} \, c_{\vec{k} \uparrow-}^\pdag - e^{+i \pi/4} \, c_{-\vec{k} \downarrow-}^\dagger\right) \text{ .} 
\end{align}\label{eq:d_f}
\end{subequations}
In the basis of Bogoliubov quasiparticles $d_{\vec{k}}$ and $f_{\vec{k}}$, the Hamiltonian $H_-$ thus takes a similar form to a normal Weyl semimetal upon replacing $m \to m \pm \Delta$.  

The subspace corresponding to $\tau^z = +1$ can be analyzed in the same way, which leads to a Hamiltonian
\begin{align}
\mathcal{H}^{\pm\Delta}_+ =&\,v_F \,\left(\hat{z}\times\vec{\sigma}\right)\cdot \vec{k} + M^{\pm\Delta}_+(k_z){} \,\sigma^z \text{ ,}\\
M^{\pm\Delta}_+(k_z) =& \left(m \pm |\Delta|\right) + \sqrt{t_S^2 + t_D^2 + 2 \, t_S t_D \cos\left(k_z d\right)}\text{ .}
\end{align}
For $m> |\Delta|$, this subspace is adiabatically connected to the topologically trivial case $m=|\Delta| = 0$,  and Weyl nodes can only appear in the $\tau^z = -1$ sector. By analogy to the normal case, we find that for $m > |\Delta|$, the spectrum of Eq.~\eqref{eq:ham_-} has up to 4 Bogoliubov Weyl nodes of pairwise opposite chiralities at $\vec{k} = \left(0,0,\pi/d\pm k_\pm^\Delta\right)^T$ with

\begin{equation}
k_\pm^\Delta = \frac{1}{d}\arccos\left(1-\frac{(m\pm|\Delta|)^2-(t_S-t_D)^2}{2 \, t_S \, t_D}\right)\label{eq:k_delta_pm}
\end{equation}
if the respective conditions

\begin{equation}
m_{c1} < m\pm|\Delta| <  m_{c2}
\end{equation}
are fulfilled. For $m\pm|\Delta| < m_{c1}$, the respective mode is adiabatically connected to the case $m=|\Delta| = 0$ and thus topologically trivial. 

If  $m< |\Delta|$, each $\vec{\tau}$ sector contains one topologically trivial mode as well as one mode that potentially has Weyl nodes. The latter now exist in the range

\begin{equation}
m_{c1} < |\Delta| \pm m<  m_{c2} \text{ ,}
\end{equation}
at the same momenta $\vec{k} = \left(0,0,\pi/d\pm k_\pm^\Delta\right)^T$ with
\begin{equation}
k_\pm^\Delta = \frac{1}{d}\arccos\left(1-\frac{(|\Delta|\pm m)^2-(t_S-t_D)^2}{2 \, t_S \, t_D}\right) \text{ ,}
\end{equation}
and the topologically trivial regime is corresponds to $|\Delta|\pm m < m_{c1}$.


\section{Characterization of the accessible phases for $\delta\Delta=0$ and $\varphi=\pi$}
\label{sec:phases}

Having recast the Weyl superconductor Hamiltonian into a more
convenient form, we will now analyze it in detail. For
  simplicity, we focus on the case $m > |\Delta|$, when all of the
  interesting physics happens in $H_-$ defined in
  Eq.~\eqref{eq:ham_-}. As the discussion in the last section implies,
  the case $m < |\Delta|$ follows upon interchanging the roles of $m$
  and $|\Delta|$ and considering different subspaces of the full
  Hamiltonian.

We first recall some results for the limiting case $|\Delta| \to 0$, in
which a normal Weyl semimetal is recovered (see
Fig.~\ref{fig:masses}(a)).\cite{Burkov_Balents_11} If the Zeeman mass
$m$ is small, $m < m_{c1}$, the system is a topologically trivial
insulator. When $m$ reaches $m_{c1}$, two Weyl nodes of opposite
chirality appear at $\vec{K} = (0,0,\pi/d)^T$. Close to these points, $\vec{k} = \vec{K}+\vec{q}$, the
dispersion is roughly given by
\begin{equation}
E \approx \pm v_F \, \vec{\sigma}\cdot\vec{q}\text{ ,}
\end{equation}
where $\pm$ defines the chirality of the node.

Upon increasing $m$, the Weyl nodes move in opposite directions along
the $\hat{k}_z$-axis and to the momenta $\vec{k} = \left(0,0,\pi/d\pm
  k_\pm^{\Delta=0}\right)^T$. For fixed $k_z$, the combined Hamiltonians
$\mathcal{H}_-^{\pm\Delta = 0}$ describe a gapped 2-dimensional Dirac
electron. The mass of the latter changes sign at the Weyl nodes. The
sign change in the Dirac mass signals a quantum Hall transition. For
small $|k_z|$, where the mass is negative, the system is still in the
topologically trivial regime. The 2-dimensional systems corresponding to
momenta outside the Weyl nodes, however, are in topologically
nontrivial quantum Hall state. Chiral surface modes appear on any
surface that is not perpendicular to $\hat{z}$ for each value of $k_z$
between the Weyl nodes. This restriction gives rise to so-called Fermi
arcs.

\subsection{Effect of pairing on nodes}
\label{sec:effect-pairing-nodes}

When superconductivity is turned on, the Hamiltonian decomposes into two
copies of itself, acting on Bogoliubov quasiparticles rather than
electrons.  This is due to the fact that superconductivity splits each
electronic state into a particle-hole symmetric and particle-hole
antisymmetric state with an energy separation $\sim 2\,|\Delta|$.  For
the Hamiltonian $H_-$, these new states correspond to $d_{\vec{k}}$ and
$f_{\vec{k}}$ defined in Eq.~\eqref{eq:d_f}.

\begin{figure}
\centering
\includegraphics[scale=0.39]{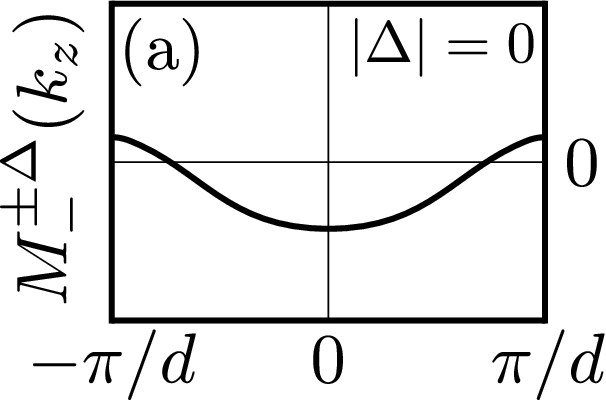}\quad~~\includegraphics[scale=0.39]{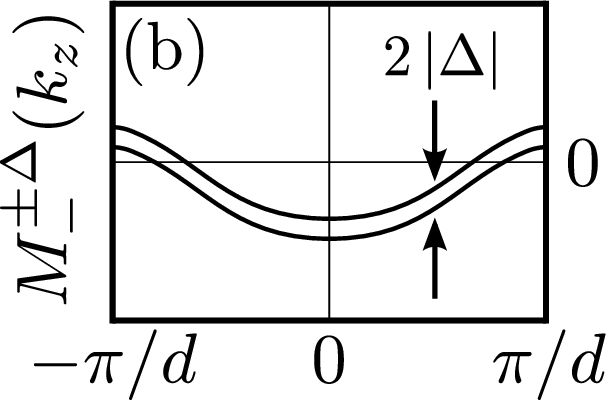}\\[0.4cm]
\includegraphics[scale=0.39]{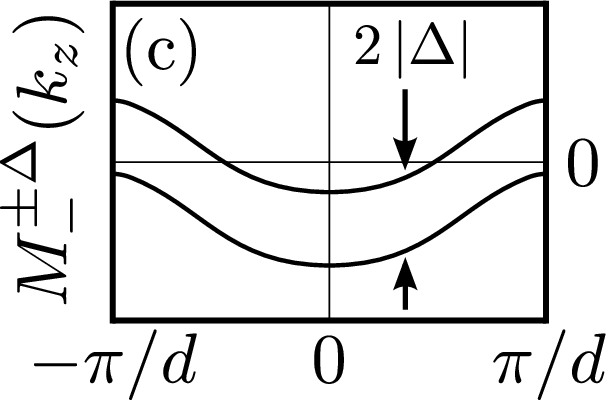}\quad~~\includegraphics[scale=0.39]{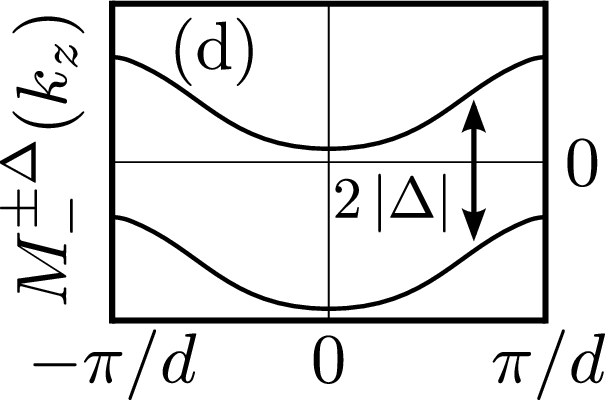}
\caption{Evolution of the masses $M_-^{+\Delta}$ (upper curve) and $M_-^{-\Delta}$ (lower curve) defined in Eq.~\eqref{eq:m_pm_delta} upon increasing $|\Delta|$. For $|\Delta| = 0$ and $m_{c1} < m < m_{c2}$, the system has two Weyl nodes of chiral electrons, located at the sign changes of $M_-^{\pm\Delta}$. With superconductivity, each Weyl nodes splits into two Bogoliubov Weyl nodes of equal chirality and opposite particle-hole symmetry. Their separation grows with increasing $|\Delta|$ from subfigures (a) to (d).}
\label{fig:masses}
\end{figure} 

Remarkably, the system does not develop a superconducting gap, but
rather each Weyl node splits into two separated Bogoliubov Weyl nodes of
opposite particle-hole symmetry (see Fig.~\ref{fig:masses}(b)). Both
Bogoliubov Weyl nodes have the same chirality, which is inherited from
the initial electronic Weyl node, and in this sense half of the
topological charge of the initial Weyl node. The particle-hole symmetric
and particle-hole antisymmetric subspaces are decoupled. For fixed
$k_z$, each subspace describes a spinless $p+i p$-superconductor, which
is known to have both a topologically trivial (``strong pairing'') and
non-trivial (``weak pairing'') phase.\cite{read_00} The transition
between the strong and weak pairing phases is still marked by the Weyl
nodes that separate trivial state at small $|k_z|$ from a non-trivial
state at large $|k_z|$.

If superconductivity is further increased, the distance between the
particle-hole symmetric and antisymmetric Weyl nodes grows. This
increases the topologically non-trivial momentum range for one of them
(the $\mathcal{H}_{-}^{+\Delta}$ subspace), and shrinks it for the other
one (the $\mathcal{H}_{-}^{-\Delta}$ subspace). For
$\mathcal{H}_{-}^{-\Delta}$, the Weyl nodes are pushed back towards
$\vec{k} = (0,0,\pi/d)^T$, where they annihilate. The corresponding
subspace is left in the topologically trivial insulating state. The Weyl
nodes of $\mathcal{H}_{-}^{+\Delta}$ move towards $\vec{k} = 0$. After
their annihilation at the origin, they leave the entire Brillouin zone
of this subspace in a topologically nontrivial insulating state. The
motion of the Weyl nodes upon increase of $|\Delta|$ is followed in
Fig.~\ref{fig:masses}. Specifically, subfigure (b) depicts the situation
where both subspaces have topologically trivial and nontrivial
momenta. Subfigure (c) corresponds to an order parameter amplitude $|\Delta|$ large enough to trivially
gap out the $\mathcal{H}_{-}^{-\Delta}$ subspace, while the
$\mathcal{H}_{-}^{+\Delta}$ subspace still has Weyl nodes. Subfigure (d)
corresponds to even larger $|\Delta|$, such that all Weyl nodes have
annihilated.

\subsection{Majorana surface states}
\label{sec:major-surf-stat}

For the topologically non-trivial momentum range of $k_z$, surface
states are expected. We model a surface perpendicular to $\hat{y}$ by
replacing $m$ and $|\Delta|$ by some smooth functions of $y$ with $m
(y), |\Delta|(y) = \text{const.}$ for $y<0$, and $m(y), |\Delta|(y) \to
0$ for $y \to +\infty$ (which realizes a trivial insulator equivalent to
the vacuum). The Hamiltonian
\begin{align}
\mathcal{H}^{+\Delta}_- =&\,v_F \,\left(k_x \sigma^y +i \frac{\partial}{\partial y} \sigma^x \right) + M^{+\Delta}_-(k_z, y) \,\sigma^z
\end{align}
indeed has eigenstates
\begin{align}
\Psi_{\text{surf}}(k_x, k_z, y) = \frac{1}{\mathcal{N}}\,e^{\int_0^y dy'\,M^{+\Delta}_-(k_z, y')/v_F}\begin{pmatrix}e^{-i\pi/4}\\e^{i\pi/4}\end{pmatrix}\label{eq:1}
\end{align}
which are normalizable and exponentially localized at the surface only
for momenta $k_z$ with $M^{+\Delta}_-(k_z) >0$ inside the sample, as
anticipated. $\mathcal{N}$ is the corresponding normalization
factor. The dispersion of the surface state is linear, $E = v_F
k_x/2$. 

The ``Majorana-ness'' of this state can be understood by counting.  In
particular, recall that the Nambu construction, Eq.~\eqref{eq:h_-},
nominally doubles the number of components of the fermionic fields.  This
implies that $\psi_k$ and $\psi_{-k}$ are not independent.  Thus
corresponding to Eq.~\eqref{eq:1} there is {\sl one} state, i.e. one
canonical (complex) fermion, for each momentum satisfying the
localization condition $M^{+\Delta}_-(k_z) >0$ with, say, $k_x >0$,
where the last condition is made to keep the states independent.
Equivalently, we can divide this complex fermion into two real ones, and
associate one real {\sl Majorana} fermion with each $k_x$, with no
restrictions on $k_x$.

In conclusion, we find that the Bogoliubov Hamiltonians
$\mathcal{H}^{+\Delta}_-$ and $\mathcal{H}^{-\Delta}_-$ essentially each
describe half of a normal Weyl semimetal.  They are subject to
respective effective Zeeman gaps $m\pm|\Delta|$. A pair of Bogoliubov
Weyl nodes of opposite chirality exists if $m_{c1} < m\pm|\Delta| <
m_{c2}$.  In the sense that two Bogoliubov Weyl nodes arise from a
single normal Weyl one, each of the former carries half of the
topological charge of the latter.  However, this notion is tied up with
the non-independence of Bogoliubov states.  The corresponding
quasiparticles are characterized by their chirality and particle-hole
symmetry rather than just chirality in the non-superconducting case. The
Bogoliubov Weyl nodes are located at momenta $\vec{k} = (0,0,k_z)^T$, with
\begin{align}
M_{-}^{\pm\Delta}(k_z) = 0 \text{ .} \label{eq:bogol_weyl_cond}
\end{align}
The vanishing of $M_{\pm}^\Delta(k_z)$ reflects a topological
transition, regarding the quasiparticles as two dimensional ones
parametrized by $k_z$. For any $k_z$ that satisfies
\begin{align}
M_{-}^{\pm\Delta}(k_z) > 0 \text{ ,} \label{eq:m_pm_cond}
\end{align}
the respective Hamiltonian $\mathcal{H}^{\pm\Delta}_-$ maps to a
topologically non-trivial spinless $p+ip$ superconductor and has a
chiral surface mode. The latter describes a Majorana particle with
linear dispersion perpendicular to $\hat{z}$ and the surface, $E = v_F
\, \vec{k}\cdot\left(\hat{e}_\perp\times\hat{z}\right)/2$. The spin of
the surface mode is locked to the direction of propagation. Negative
values of $M_{-}^{\pm\Delta}(k_z)$, on the contrary, correspond to a
trivial insulator.

\subsection{Topological phase diagram}
\label{sec:topol-phase-diagr}

\begin{figure}
\centering
\includegraphics[scale=0.75]{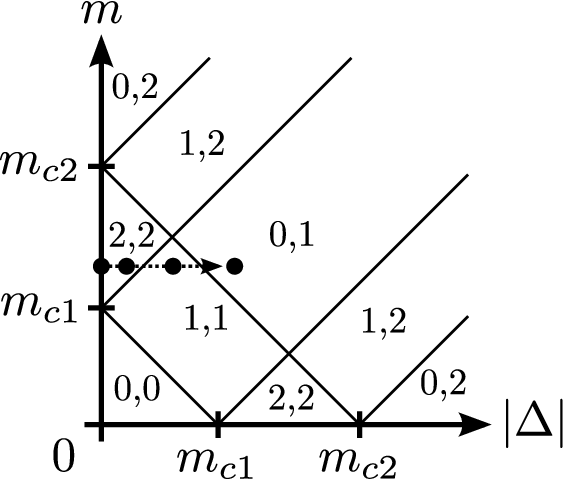}
\caption{Phase diagram of a Weyl superconductor in a TI/SC heterostructure as a function of Zeeman gap $m$ and proximity induced superconducting order parameter amplitude $|\Delta|$. The values of $m_{c1}$ and $m_{c2}$ are set by the tunneling amplitudes between the surface Dirac layers of the heterostructure depicted in Fig.~\ref{fig:heterostructure}, see Eq.~\eqref{eq:mc_i}. Each phase is characterized by $n_b$, the number of pairs of bulk Bogoliubov Weyl nodes, and $n_s$, the number of two-dimensional Majorana surface modes. The phases are labeled as $n_b,n_s$. The black dots locate the different subfigures of Fig.~\ref{fig:masses}.}
\label{fig:phase_diag}
\end{figure} 

The Hamiltonians $\mathcal{H}^{\pm\Delta}_-$ can separately be tuned
from a topologically trivial to topologically non-trivial state by
changing both $m$ and $|\Delta|$. In our model, these two parameters can be tuned separately, although a finite magnetization can in principle affect the proximity induced superconductivity in the TI/SC interfaces. An analysis of the effect of the magnetization on the gap (and of the superconductivity on the magnetic ordering) requires a theory of the superconducting mechanism, like BCS theory, which goes well beyond the treatment here and we think deserves a separate study from this manuscript.

If we exclude a substantial magnetic field due to magnetic impurities, they could most importantly affect the superconductivity in the TI/SC interfaces by an exchange coupling of the nearby superconductor layers to the magnetic impurities. A weakening of the superconductivity there would in turn diminish the proximity effect at the interfaces. Since however the superconductor layers are separated from the magnetic impurities by the interfaces, the exchange coupling of the SC to the magnetic impurities is certainly much weaker than the Zeeman term in the interfaces. It is therefore reasonable to neglect the explicit dependence of $|\Delta|$ from $m$. Similarly, a proximity effect of the SC on the impurities potentially weakening their ferromagnetic ordering is disregarded. This approximation would break down if either the magnetic field of the impurities was important, or if the tunnel coupling between the superconductor and the magnetic impurities was strong.

Neglecting their weak interdependence, a simultaneous modification of $m$ and $|\Delta|$ allows to access a number of
different phases with $n_b = 0, 1, 2$ pairs of Bogoliubov Weyl nodes in
the bulk and $n_s=0, 1, 2$ two-dimensional Majorana surface modes
(potentially living in a restricted
$k_z$-range). Fig.~\ref{fig:phase_diag} shows the phase diagram as a
function of $m$ and $|\Delta|$, the phases are labeled according to
their values of  ($n_b,n_s$). The phase diagram is mirror symmetric for
negative values of $m$. The tunability of Weyl superconductors (which
can be mapped to four copies of a spinless $p+i p$ superconductor per
value of $k_z$, only two of which are potentially topologically
non-trivial) is similar to the one of superconducting quantum anomalous
Hall insulators (which can be mapped to two topologically potentially
non-trivial copies of a spinless $p+ip$ superconductor), although the
increased complexity leads to a richer phase diagram for Weyl
superconductors.\cite{qi_10} Of particular interest are the phases
$(0,1)$, which corresponds to a truly topological superconductor (in class
D and thus with two-dimensional topological invariants similarly to a three-dimensional quantum anomalous Hall state), and $(1,1)$, which is
precisely half of a normal Weyl semimetal.  We will analyze them further
in the next section.\cite{Schnyder_08}


\section{Vortices in Weyl superconductors}
\label{sec:vortices}

One of the most interesting features of two dimensional topological
superconductors is that they may host a zero energy Majorana mode
localized around a vortex.  Collections of such Majorana bound states
allow for non-local storage of quantum information, which may make the
stored information less sensitive to decoherence.\cite{Kitaev_01}
This motivates the analysis of vortices in Weyl superconductors.  We
specifically discuss the behavior of vortices in the simplest $(0,1)$
and $(1,1)$ phases of Fig.~\ref{fig:phase_diag}, which minimize the
number of surface Majorana modes.  The vortex physics in other phases is
qualitatively similar, but may involve more Majorana modes.

The suppression of superconductivity inside a vortex puts its core in
either the $(0,2)$, $(2,2)$ or $(0,0)$ phase. For simplicity, we
consider the core to be is in the trivially insulating $(0,0)$ phase.
In general, the finite size of the core suggests that in any case the
core cannot be sharply distinguished from a trivial state, in a full
treatment.  The $(0,0)$ state can always be realized for an appropriate
choice of $m$ and $|\Delta|$.   Nevertheless, our results are not affected
by this assumption. Different values of $m$ and $|\Delta|$ will at most change the number and/or direction of propagation of the interface Majorana modes.  

\begin{figure}
\centering
\includegraphics[scale=0.1]{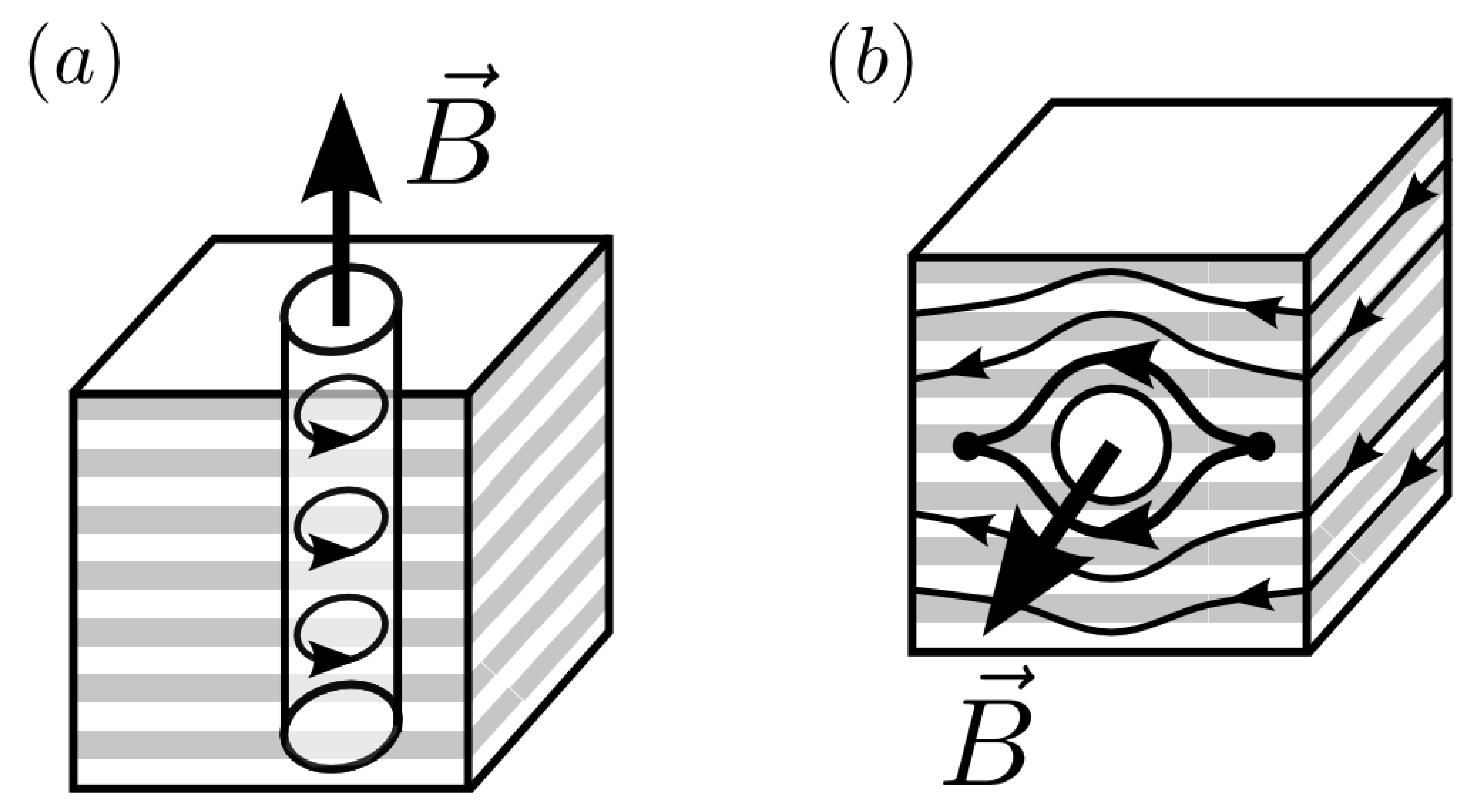}
\caption{The two classes of vortices in Weyl superconductors. Subfigure (a) sketches a vortex along the superlattice axis $\hat{z}$, with bound states along a tube through the whole sample. Subfigure (b) depicts a vortex perpendicular to $\hat{z}$. Whereas there are no states bound to the vortex, the surface states can be used for Majorana interferometry (thick line).}
\label{fig:vortex}
\end{figure}

Because the heterostructure has one special direction, namely the
$\hat{z}$-axis along which the different layers are stacked, vortices
parallel and perpendicular to $\hat{z}$ have to be distinguished, as
depicted in Fig.~\ref{fig:vortex}.  The qualitative physics of vortices
can be understood by analogy to ${}^3$He-A, which also is a Weyl
superconductor.\cite{volovik_book, volovik_11} As pointed out in
Ref.~\onlinecite{volovik_11}, the momentum range in which Majorana bound
states at a vortex exist is proportional to $\hat{e}_v\cdot\hat{z}$,
where $\hat{e}_v$ is the direction of the vortex. In particular, a
vortex perpendicular to $\hat{z}$ has no bound states. In the following,
we will analyze the behavior of vortices in TI/SC heterostructures in
more detail.

\subsection{Vortex along the superlattice axis}

At first, we turn to a magnetic field $\vec{B}$ applied along $\hat{z}$,
the stacking axis of the heterostructure. For modest field strengths,
only few vortices are present, and interactions between vortices can be
neglected. This situation is sketched in Fig.~\ref{fig:vortex}(a). By
assumption, the vortex core is in a topologically trivial insulating
state. The boundary of the vortex is thus equivalent to an interface
between a Weyl superconductor and vacuum and has 1 Majorana edge
mode. If the Weyl superconductor is in the $(1,1)$ phase, this mode has a
restricted range of momenta $k_z$ (it lives "between the Bogoliubov Weyl
nodes"), the $(0,1)$ phase has interface modes for any momentum $k_z$. We
restrict the discussion to $|\Delta| < m$ when all relevant physics
happens in $H_-$, but the results can easily be generalized.

Exploiting the cylindrical symmetry with respect to the vortex axis, we
model the latter by a radially dependent Zeeman gap $m$ while (restoring the global superconducting phase) the 
superconducting order parameter $|\Delta|$ is replaced by $|\Delta(r)|e^{i\varphi(\phi)}$, where
$\varphi(\phi) = \varphi_0-(\Phi/\Phi_0)\,\phi$ is the phase of the order parameter. The latter is now twisted due to the presence of a magnetic field ($\phi$ denotes the angular coordinate). The twist is proportional to the flux $\Phi$ trapped by the vortex, which itself is quantized in units of $\Phi_0 = hc/2e = \pi/e$, as usual for superconductors. The phase $\varphi_0$ corresponds to the superconducting phase without magnetic field. The radius of the vortex is
considered to be $R$, and $m(r)$ and $|\Delta(r)|$ are smooth functions
interpolating between fixed values $m$ and $|\Delta|$ for $r>R$, and
$|m| < m_{c1}, |\Delta| = 0$ inside the core of the vortex. The magnetic
field is $\vec{B} = B\,\hat{z}$ inside the vortex and for simplicity assumed to vanish
everywhere else. This gives rise to a vector
potential
\begin{align}
\vec{A}(\vec{r}) &= A(r)\, \hat{e}_\phi \text{ ,}\\
A(r) &= \frac{B r}{2} \,\Theta(R-r) + \frac{B R^2}{2r} \,\Theta(r-R)\nonumber
\end{align}
in $\hat{e}_\phi$ direction, that is taken into account by minimal
coupling $\vec{k} \to \vec{k}-e\vec{A}$ in the Hamiltonian
\eqref{eq:heterostructure_ham1}. After a canonical transformation $\sigma^x \to
-\sigma^y, \sigma^y \to \sigma^x$, the relevant Hamiltonian $H_-$ in
Eq.~\eqref{eq:h_-} becomes
\begin{subequations}
\begin{align}
H_-&= \sum_{k_z}\int d^2r\,\psi_{k_z}^\dagger(\vec{r})\, \mathcal{H}_- \,\psi_{k_z}^\pdag(\vec{r})\text{ ,}\\[0.2cm]
\mathcal{H}_- &= \begin{pmatrix}\mathcal{H}_A
  &|\Delta(r)|\,e^{i\varphi(\phi)}\,\sigma^z
  \\|\Delta(r)|\,e^{-i\varphi(\phi)}\,\sigma^z&\mathcal{H}_{-A} \end{pmatrix}\text{
  ,}\\[0.2cm] 
\mathcal{H}_A &=M_-(k_z, r) \,\sigma^z + v_F \begin{pmatrix}0&-i \, e^{-i \phi}\\ -i \, e^{i \phi}& 0\end{pmatrix}\frac{\partial}{\partial r}\\
&+ v_F \begin{pmatrix}0&-\, e^{-i \phi}\\ e^{i \phi}& 0\end{pmatrix}\left(\frac{1}{r}\frac{\partial}{\partial \phi}+i e\, A(r)\right)\nonumber \text{ .}
\end{align}\label{eq:sc_vortex_ham}
\end{subequations}
For any given $k_z$, this Hamiltonian may be interpreted as two copies of a spinless $p+ip$ superconductor threaded by a magnetic flux. By analogy, the vortex binds one Majorana zero mode per topological value
of $k_z$ and per topological subsector if it traps an odd number of flux quanta, and no zero mode for an even number
of trapped flux quanta.\cite{read_00} Assuming that there is only a single topologically non-trivial subsector, one can thus define a unique zero energy Majorana mode bound to the vortex. See appendix \ref{append:vortex_state} for more details.

Physically, the Majorana bound state can be understood in terms of an
Aharonov-Bohm like phase, a Berry phase and a geometrical phase for the
Majorana surface states.  Consider the topologically equivalent
situation of a Weyl superconductor with a tube-like hole along the
$\hat{z}$ axis. Without a magnetic field inside the hole, we know that
chiral Majorana surface states exist when $k_z$ is chosen in the range
where the two-dimensional superconductor is in the topological phase. Since the spin
is locked to the momentum, the surface states pick up a Berry phase of
$\pi$ upon encircling the hole once. This shifts the zero momentum mode
away from zero energy and can be interpreted as effectively antiperiodic
boundary condition on the geometrical phase in order to counterbalance
the Berry phase.  If now a unit flux is threaded through the tube-like
hole, the surface states pick up an additional phase of
$\pi$. The latter derives from the winding of the order parameter phase, and is similar to an Aharonov-Bohm effect. It compensates the Berry phase and thus allows for zero energy bound states.  Similar effects have also been discussed for confined magnetic
flux tubes imposed in 3-dimensional strong topological
insulators.\cite{rosenberg_10, chiu_11} For momenta $k_z$ which are in
the topologically trivial range, of course, no bound states exist both with and without
magnetic flux. Because the magnetic field vanishes outside the vortex,
the topological character and especially the existence of surface states
is unchanged there. We thus conclude that a vortex with an odd number of
flux quanta traps a Majorana zero mode for every topologically
non-trivial value of $k_z$.

In a more realistic model, the Majorana bound states do not form
totally flat bands as a function of $k_z$. The presence of a zero energy
Majorana mode for odd-integer fluxes is however partially robust. As an
effective model at lowest energies, we consider the zero energy band of
Majorana modes as a function of $k_z$. After
transforming to Wannier orbitals, we obtain a set of Majorana bound
states at different heights $z$, as depicted in
Fig.~\ref{fig:vortex}(a). This Hamiltonian can be interpreted as a
1-dimensional chain of decoupled sites. Next, we introduce a small
hopping along the chain, thus allowing the Majoranas to move up and down
the vortex tube. In dimensionless units, their dispersion is given by
\begin{equation}
E = -\cos(k_z) \text{ .}
\end{equation}
Therefore, zero energy Majorana bound states exist if
\begin{enumerate} 
\item a Majorana bound state can be defined for $k_z=\pm\pi/2$, i.e. $M^{\pm\Delta}_-(\pm \pi/2) >0$, and if
\item $k_z$ can take the values $\pm\pi/2$.
\end{enumerate}
The first condition is always fulfilled in the $(0,1)$ phase, but
depends on the exact position of the Bogoliubov Weyl nodes in the
$(1,1)$ phase. The second condition depends on the number of layers of
the TI/SC heterostructre and the boundary conditions. Choosing for
instance hard wall boundary conditions, one finds exactly one zero
energy Majorana mode if the system has an odd number of superlattice
layers, and no zero energy Majorana modes for an even number of
layers.  This result is quite natural, in a weak tunneling picture.
Majorana states in a pair of layers can mix to form a Dirac fermion,
moving away from zero energy.  Only for an odd number of layers is an
unpaired Majorana left behind at zero energy.

\subsection{Vortex perpendicular to the superlattice axis}

Viewing the vortex, as in the previous subsection, as a cylindrical
hole enclosing a flux, the discussion in
Sec.~\ref{sec:major-surf-stat} implies that a vortex perpendicular to
the axis $\hat{z}$ of the heterostructure should also host Majorana
modes.  In our model, they run between the front and back surfaces of
the heterostructure on the side walls of the vortex. For a thin vortex, however, already a small coupling across the
flux line is sufficient to hybridize and consequently gap out these two states. We thus recover the result of
Ref.~\onlinecite{volovik_11}.

The hole comprising the vortex also introduces a new edge at the
sample boundary (the circular ends of the cylindrical hole). The
nearby surface states will rearrange in order to host the vortex and
locally run along this new edge, as depicted in
Fig.~\ref{fig:vortex}(b). While there are no states bound to the
vortex, a special class of surface state paths allows for Majorana
interferometry, depicted by thick lines in
Fig.~\ref{fig:vortex}(b).  It may be interesting to study experimental
measurements of interference in such structures.


\section{Experimental probes}
\label{sec:exp}
\subsection{Anomalous thermal Hall effet}

Because Majorana particles do not carry electric charge, a natural
way to measure surface Majorana states is to detect their thermal
transport. We again focus on a Weyl superconductor in the $(0,1)$ or
$(1,1)$ phase. As there is only a single Majorana surface modes, these
phases should exhibit only half of the thermal transport of a normal
Weyl semimetal in the corresponding regime.

On a surface perpendicular to $\hat{y}$, thermodynamics can be calculated from the effective Majorana surface partition function
\begin{align}
Z &= \int \mathcal{D}\left(\overline{\Psi}_{\omega_n,k_x,k_z}, \Psi_{\omega_n,k_x,k_z}\right)_{k_x>0} e^{-\mathcal{S}} \text{ ,}\\
\mathcal{S} &= \sum_{\substack{\omega_n,\\k_x>0, k_z}} \overline{\Psi}_{\omega_n,k_x,k_z}\left(-i\,\omega_n + v_F\,k_x\right) \Psi_{\omega_n,k_x,k_z} \text{ ,}
\end{align}
where the operators $\Psi_{\omega_n,k_x,k_z}^\dagger$ create
excitations above the Bogoliubov vacuum.  Note the restriction to
$k_x>0$, which is because pairs of Majorana fermions at $k_x$ and
$-k_x$ have been recombined into the canonical $\Psi$ fermion
(c.f. Sec.~\ref{sec:major-surf-stat}).

\begin{figure}
\centering
\includegraphics[scale=0.6]{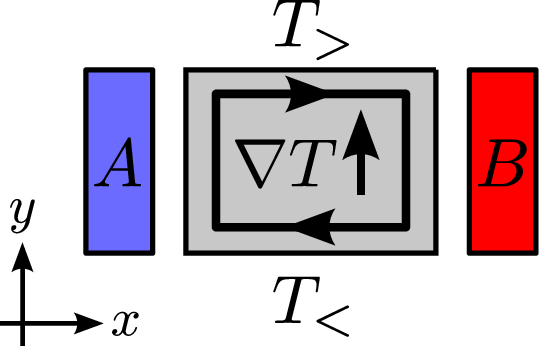}
\caption{Sketch of the anomalous thermal Hall effect in Weyl superconductors. The sample is shown from above. A thermal gradient $\nabla T$ is applied across the sample. The upper surface is at a temperature $T_>$ larger than the temperature $T_<$ of the lower surface. This leads to a net heat current from side $A$ to side $B$ perpendicular to the temperature gradient transported by the surface modes running around the sample.}
\label{fig:thermal_hall}
\end{figure} 

If a thermal gradient $\nabla T$ is applied across the Weyl semimetal,
each surface mode transports heat only in its direction of
propagation. Therefore, the thermal gradient leads to a net heat
transport perpendicular to $\nabla T$, as depicted in
Fig.~\ref{fig:thermal_hall}. This phenomenon is known as the thermal
Hall effect. It has been proposed as an experimental
signature of various other chiral edge states, for example in the spin
Hall effect, the fractional quantum Hall effect or topological
superconductors. \cite{kane_97, senthil_99, read_00, wang_11}  We presume any bulk transport to be
parallel to the gradient, so that it can be separated from the surface
contribution.  In any case, the dependence upon field, density,
etc. of any possible bulk contribution would be very different from
that of the surface one.

For concreteness, consider a temperature gradient $\nabla T$ imposed
across the sample in the $\hat{y}$ direction.  This leads to a net
difference in the distribution of quasiparticles on the $y=0$ and
$y=L_y$ surfaces.  The result is an excess heat current $I_Q$, in the
$x$ direction, which defines the thermal Hall conductance $K_{xy}$,
according to
\begin{align}
I_{Q} = K_{xy}\, |\nabla T| \text{ .}
\end{align}
For small temperature differences between the surfaces, the excess heat current is obtain by
differentiation, and we obtain
\begin{align}
 K_{xy} &= \sum_{k_z} \, \int_0^{\infty} \frac{d k_x}{2\pi} \, v_F^2 k_x\, \frac{\partial \, n_F(v_F k_x)}{\partial T} \text{ ,} \\
 &=  \sum_{k_z} \frac{1}{2} \frac{k_B^2\pi^2 T}{3 h},
\end{align}
with $k_z$ being summed over all topologically non-trivial values for
the given phase of the Weyl superconductor (either $(0,1)$ or $(1,1)$)
and $n_F$ denoting the Fermi-Dirac distribution at the temperature $T$. Note that we have restored physical units such as Boltzmann's
constant $k_B$ and Planck's constant $h$ for concreteness. As expected, the surface of
a Weyl superconductor has half of the thermal Hall conductance of a
quantum Hall edge state per allowed momentum $k_z$. This is not
surprising because the thermal Hall coefficient is proportional to the
central charge $c$ of the surface modes, $K_{xy} = c \, \pi^2
k_B^2 T / (3 h)$, similar to the heat
capacity.\cite{read_00,boethe_affleck_86}

Coming back to the Weyl superconductor in the $(0,1)$ or $(1,1)$ phase,
the thermal Hall effect has an anomalous coefficient proportional to
the distance $2 k_+^\Delta$ between the Weyl nodes defined in
Eq.~\eqref{eq:k_delta_pm}.  Concretely, the thermal Hall conductance
is proportional to the length of the system in the $\hat{z}$
direction, $K_{xy}=\kappa_{xy} L_z$, with
\begin{equation}
 \kappa_{xy} = \frac{1}{2} \frac{k_B^2\pi^2 T}{3 h}\, \frac{k_+^\Delta}{\pi} \text{ .}
\end{equation}

In the $(0,1)$ phase, where $k_+^\Delta = \pi/d$, each TI layer
contributes the full Majorana quantum $(1/2)\, \pi^2 k_B^2 T / (3 h)$
to the thermal Hall coefficient. Although thermal transport
measurements are experimentally demanding, the higher dimensionality
of the surface states in a Weyl superconductor as compared to
fractional or spin quantum Hall edge states hopefully tends to result
in more approachable experiments.

\subsection{Electrical transport}

As discussed in Sec.~\ref{sec:phases}, the surface physics of a Weyl
superconductor can be understood as layers of spinless $p+ip$
superconductors stacked in momentum space along $k_z$, with potentially associated edge states. In order to minimize bulk transport, we now specialize to the $(0,1)$ phase. The surface of
the Weyl superconductor is then equivalent to just one non-trivial spinless
$p+ip$ superconductor edge state per value of $k_z$. In this phase,
electric transport experiments that have been proposed for $p+ip$
superconductors can simply be transferred to Weyl superconductors. The
general idea is to bring different samples with Majorana edge modes
into contact. Whenever an interface has two edge modes running into
the same direction, electrons can tunnel into the interface by
decomposition in the two Majorana particles. These two Majorana
particles can then be transported in parallel, giving rise to a
one-directional electronic transport channel along the
interface.\cite{Serban_10} In alternative setups, the two Majoranas
can be separated and recombined with different Majorana modes, which
leads to distinct signatures in conductance and noise.\cite{chung_11}
The latter experiments are however less appropriate for Weyl
superconductors where each surface has a large number of generically
coupled Majorana modes at different values of $k_z$. 


\section{Summary and conclusions}

We have shown how a variety of gapless and/or topological
superconducting phases can be achieved in superconducting--topological
insulator superlattices.  These phases are analogous to quasiparticle
states of ${}^3$He, The most interesting $(0,1)$ and $(1,1)$ phases
exhibit Majorana surface states on some surfaces, and bound to the
cores of vortices.  Particularly in the gapless phases, such as
$(1,1)$, these Majorana states exist only for a range of momenta,
$k_z$, along the modulated direction of the superlattice.  In such a
case, no local (in $z$) description of the Majorana modes is possible,
as opposed to the situation in the $(0,1)$ phase, in which the
Majorana modes can be modeled in terms of a real-space tight-binding
Hamiltonian in the $z$ direction, and the state can be considered as a
sort of stack of two-dimensional topological superconductors.

It is hoped that the proposed structures might be explored
experimentally in the future.  While we do not discuss materials in any
detail here, we note that recent studies have shown that
Cu$_x$Bi$_2$Se$_3$\cite{zhang_09, peng_09, zhang_10} becomes a
superconductor with $x\approx 0.14$, while it is a topological insulator
for $x=0$, so that a superlattice with modulated $x$ might be a
candidate realization of this proposal.  Some {\sl ab initio} modeling
of such a superlattice would probably be useful prior to any
experimental attempts. Alternatively, spin-triplet superconductors have recently been identified to exhibit Weyl superconducting phases as well.\cite{Sau_11, Wu_12}

There is significant scope for further theoretical study of Weyl and
topological superconductors in three dimensions.  The Majorana surface
states of $(0,1)$ and $(1,1)$ phases are rather analogous to the
``chiral surface sheaths'' which occur in three-dimensional quantum
Hall systems,\cite{Chalker_95,Balents_96}
where interesting vertical transport, quantum interference, and
universal conductance fluctuations have been studied, and it would be
interesting to see how such phenomena translate to the superconducting
case.  We have also not touched on the Adler-Bell-Jackiw anomaly
associated with Weyl points.  This has been discussed recently for
normal Weyl semimetals, where it may lead to anomalous
magnetotransport.\cite{Aji_11}   It is not obvious what the consequences are for
Weyl superconductors.  One might also consider Josephson effects for
currents along the $z$ axis.  We leave these questions for future work.

\begin{acknowledgments}
  We acknowledge discussions with Victor Galitski, and Xiaoliang Qi. L.B. was supported by NSF grants DMR0804564{, DMR1506119, and} PHY05-51164.
  T.M. gratefully acknowledges the hospitality of KITP, where part of
  this work was done, as well as financial support by SFB 608{, SFB 1143, and} FOR 960 of the {DFG}, and the Bonn-Cologne graduate school (BCGS).
\end{acknowledgments}

\appendix

\section{Gaplessness of inversion symmetric Weyl superconductors}
\label{append:gap_or_no_gap}
In this appendix, we discuss the qualitative behavior of Weyl semimetals under superconducting proximity effect. We assume that either time reversal or inversion symmetry is conserved, while the respective other symmetry needs to be broken for the system to exhibit Weyl physics. As has been mentioned in the introduction, the fate of the Weyl superconductor depends on which of the symmetries is conserved. For inversion symmetric, time reversal symmetry broken Weyl semimetals, the presence of a Weyl node at $\vec{k_0}$ implies the presence of a Weyl node of {\it opposite} chirality at $-\vec{k_0}$. In Weyl semimetals with broken inversion symmetry, however, time reversal symmetry guarantees the presence of Weyl nodes of {\it equal} chirality at $\pm \vec{k}_0$. When a superconducting proximity effect is turned on, the low energy modes at these two Weyl nodes mix. However,
the superconducting correlations more precisely couple electrons on one Weyl node to
holes on the other, instead of electrons to electrons as a more standard perturbation
would do. This effectively inverts the chirality of one of the Weyl nodes. Consequently, the proximity effect in inversion symmetric systems {\it effectively} mixes Weyl nodes of the {\it same} chirality, and no gap opens. In time reversal symmetric systems, on the contrary, the mixed Weyl nodes effectively have {\it opposite} chiralities, and a gap is to be expected.

To be more concrete, we consider the effective low energy theory of a Weyl semimetal, which corresponds to electrons living close to Weyl nodes. Since standard superconductivity couples electrons at momenta $\pm \vec{k}$, we focus our effective model on two of the Weyl nodes located at momenta $\pm \vec{k}_0$. For an inversion symmetric system, where the nodes are of opposite chirality, this already describes a complete minimal model. For a time reversal symmetric system, the nodes at $\pm \vec{k}_0$ have the same chirality, and there must exist at least two additional nodes, say at $\pm\vec{k}_1$, of the respective opposite chirality. Since these two pairs of nodes are decoupled, we can understand the system as two copies of the following Hamiltonian \eqref{eq:append_h_delta} describing only two Weyl nodes. The latter thus allows us to decide on the presence or absence of a gap.

As advertized, our result will only depend on whether the two initial Weyl nodes have the same or opposite chirality. We therefore assume one Weyl node to have positive chirality, $H_1 \sim \vec{\sigma}\cdot \vec{k}$, while the second node is so far keep in a general notation, $H_2 \sim \pm\,\vec{\sigma}\cdot \vec{k}$ in order to tackle both the time reversal symmetric and inversion symmetric cases simultaneously. The electrons close to these nodes are described by the operators $c_{1,\vec{k},\sigma}$ and $c_{2,\vec{k},\sigma}$, respectively. We furthermore measure the momenta relative to the respective Weyl nodes, such that the non-superconducting Hamiltonian reads

\begin{align}
 H_0 &= \sum_{\vec{k}}\,(c_{1,\vec{k},\uparrow}^\dagger, c_{1,\vec{k},\downarrow}^\dagger)\,\left( v_F\,\vec{\sigma}\cdot\vec{k}\right)\,\begin{pmatrix} c_{1,\vec{k},\uparrow}^\pdag\\ c_{1,\vec{k},\downarrow}^\pdag \end{pmatrix}\label{eq:append_h0} \\
&+ \sum_{\vec{k}}\,(c_{2,\vec{k},\uparrow}^\dagger, c_{2,\vec{k},\downarrow}^\dagger)\,\left(\pm\,v_F\, \vec{\sigma}\cdot\vec{k}\right)\,\begin{pmatrix} c_{2,\vec{k},\uparrow}^\pdag\\ c_{2,\vec{k},\downarrow}^\pdag \end{pmatrix}\nonumber\\
&= \sum_{\vec{k}}\,\Psi_{\vec{k},0}^\dagger\begin{pmatrix}v_F\,\vec{\sigma}\cdot\vec{k}&0\\0&\pm\,v_F\,\vec{\sigma}\cdot\vec{k}\end{pmatrix}\,\Psi_{\vec{k},0}^\pdag\nonumber
\end{align}
with $\Psi_{\vec{k},0} = (c_{1,\vec{k},\uparrow}, c_{1,\vec{k},\downarrow}, c_{2,\vec{k},\uparrow}, c_{2,\vec{k},\downarrow})^T$. We assume that the superconducting part of the Hamiltonian only contains terms of the from

\begin{equation}
 H_{SC} \sim h(\vec{k})\,c_{1,\vec{k},\sigma}^\dagger c_{2,-\vec{k},\sigma'}^\dagger + \text{h.c.} \text{ ,}
\end{equation}
which in particular includes s-wave and p-wave pairing. It is now useful to rewrite the non-superconducting part of the Hamiltonian as

\begin{equation}
 H_0= \sum_{\vec{k}}\Psi_{\vec{k},\Delta}^\dagger\,\begin{pmatrix}v_F\,\vec{\sigma}\cdot\vec{k}&0\\0&\mp\,v_F\,\vec{\sigma}\cdot\vec{k}\end{pmatrix}\,\Psi_{\vec{k},\Delta}^\pdag\text{ ,}
\end{equation}
where $\Psi_{\vec{k},\Delta} = (c_{1,\vec{k},\uparrow}^\pdag, c_{1,\vec{k},\downarrow}^\pdag, c_{2,-\vec{k},\downarrow}^\dagger, -c_{2,-\vec{k},\uparrow}^\dagger)^T$. We note that the sign of the second Weyl node in the Hamiltonian has to be reversed due to the inversion of creation and annihilation operators. This precisely corresponds to the effective inversion of the chirality of the second node under proximity effect, see above. Including a general superconducting term, the full Hamiltonian $H = H_0 + H_{SC}$ can be written as
\begin{align}
 H&= \sum_{\vec{k}}\,\Psi_{\vec{k},\Delta}^\dagger\,\mathcal{H}\,\Psi_{\vec{k},\Delta}^\pdag\text{ ,}\label{eq:append_h_delta}\\
\mathcal{H}&=\begin{pmatrix}v_F\,\vec{\sigma}\cdot\vec{k}&\left[\alpha(\vec{k})\,\mathds{1}_{\sigma} + \vec{\beta}(\vec{k})\cdot\vec{\sigma}\right]\\\left[\alpha(\vec{k})^*\,\mathds{1}_{\sigma} + \vec{\beta}(\vec{k})^*\cdot\vec{\sigma}\right]&\mp\,v_F\,\vec{\sigma}\cdot\vec{k}\end{pmatrix}\text{ .}
\end{align}
For an inversion symmetric system, where the lower (plus) sign applies, the diagonal is proportional to the unit matrix in Nambu space. Superconductivity can therefore never open up a gap, but only shift the Bogoliobov Weyl nodes in momentum space. For time reversal symmetric systems, the Hamiltonian is however generically gapped. As an example, we consider s-wave superconductivity. The latter corresponds to

\begin{align}
 H_{\rm s-wave} &= \sum_{\vec{k}} \Delta\, c_{1,\vec{k},\uparrow}^\dagger c_{2,-\vec{k},\downarrow}^\dagger + \text{ h.c.} \\
&= \sum_{\vec{k}} \frac{\Delta}{2}\, \left(c_{1,\vec{k},\uparrow}^\dagger c_{2,-\vec{k},\downarrow}^\dagger-c_{2,\vec{k},\downarrow}^\dagger c_{1,-\vec{k},\uparrow}^\dagger\right)  + \text{ h.c.}\text{ ,}\nonumber
\end{align}
where we neglect the superconducting phase for simplicity, i.e.~$\Delta = |\Delta|$. The total Hamiltonian then reads
\begin{equation}
 H= \sum_{\vec{k}}\,\Psi_{\vec{k},\Delta}^\dagger\,\begin{pmatrix}v_F\,\vec{\sigma}\cdot\vec{k}&\frac{\Delta}{2}\,\mathds{1}_{\sigma}\\\frac{\Delta}{2}\,\mathds{1}_{\sigma}&-\,v_F\,\vec{\sigma}\cdot\vec{k}\end{pmatrix}\,\Psi_{\vec{k},\Delta}^\pdag\text{ .}
\end{equation}
The eigenvalues of this Hamiltonian are easily found as

\begin{equation}
 E = \pm\sqrt{(v_F\,\vec{\sigma}\cdot\vec{k})^2+\frac{|\Delta|^2}{4}} \text{ ,}
\end{equation}
and the system is gapped as expected. Finally, consider adiabatically restoring inversion symmetry. Throughout this process, superconductivity ensures the system to be gapped. A time reversal symmetric, inversion symmetry broken Weyl superconductor can thus adiabatically be connected to the trivial state respecting both symmetries, and is therefore a trivial insulator itself.

\section{Expression of the zero energy Majorana bound state}
\label{append:vortex_state}
A vortex in a Weyl superconductor traps a unique zero energy bound state if it contains an odd number of flux quanta. To explicitly show this, let us first discuss the bound state for a simple limiting case where the algebra can be done explicitly, and then turn to the general solution.

The limiting case is defined as follows. We assume that the Zeeman mass $m$ is constant in the entire Weyl superconductor (and in particular takes the same value inside and outside the vortex). Moreover, we assume that $m_{c_1} < m < m_{c_2}$, such that there is one momentum $k_z = k_z^0$ with $M_-(r,k_z^0) = 0$ everywhere. As follows from Fig.~\ref{fig:phase_diag}, we are then able to find a $|\Delta| = \Delta_0$ outside the vortex such that only one subsector is topologically non-trivial, and expect a single zero energy Majorana mode bound to the vortex for this choice of $|\Delta|$. In addition, we assume that there is only a single flux quantum inside the vortex. Outside the vortex, the Hamiltonian \eqref{eq:sc_vortex_ham} reads for $k_z = k_z^0$

\begin{subequations}
\begin{align}
H_-&= \int_{r>R} d^2r\,\psi_{k_z^0}^\dagger(\vec{r})\, \mathcal{H}_- \,\psi_{k_z^0}^\pdag(\vec{r})\text{ ,}\\[0.2cm]
\mathcal{H}_- &= \begin{pmatrix}\mathcal{H}_B
  &|\Delta(r)|\,e^{i\varphi(\phi)}\,\sigma^z
  \\|\Delta(r)|\,e^{-i\varphi(\phi)}\,\sigma^z&\mathcal{H}_{-B} \end{pmatrix}\text{
  ,}\\[0.2cm] 
\mathcal{H}_B &=v_F \begin{pmatrix}0&-i \, e^{-i \phi}\\ -i \, e^{i \phi}& 0\end{pmatrix}\frac{\partial}{\partial r}\\
&+ v_F \begin{pmatrix}0&-\, e^{-i \phi}\\ e^{i \phi}& 0\end{pmatrix}\left(\frac{1}{r}\frac{\partial}{\partial \phi}+i e\frac{B R^2}{2r}\right)\nonumber \text{ .}
\end{align}\label{eq:sc_vortex_ham_append}
\end{subequations}
The order parameter amplitude $|\Delta(r)|$ goes to zero in the vortex core and takes the value $|\Delta(r)| = \Delta_0$ far away from the vortex. For $k_z = k_z^0$, this Hamiltonian has two linearly independent normalizable zero energy bound state solutions,

\begin{align}
\Psi_1^{\rm outer} &= \frac{1}{\mathcal{N}''}\,\frac{1}{\sqrt{r}}\,e^{-\int_R^r dr'\,|\Delta(k_z^0, r')|/v_F}\begin{pmatrix}e^{-i\phi}\\0\\0\\i\,e^{i\phi}\end{pmatrix} \text{ ,}\\
\Psi_2^{\rm outer} &= \frac{1}{\mathcal{N}''}\,\frac{1}{\sqrt{r}}\,e^{-\int_R^r dr'\,|\Delta(k_z^0, r')|/v_F}\begin{pmatrix}0\\i\\1\\0\end{pmatrix} \text{ .}
\end{align}
Inside the vortex. i.e.~for $r<R$, where

\begin{align}
\mathcal{H}_B &=v_F \begin{pmatrix}0&-i \, e^{-i \phi}\\ -i \, e^{i \phi}& 0\end{pmatrix}\frac{\partial}{\partial r}\\
&+ v_F \begin{pmatrix}0&-\, e^{-i \phi}\\ e^{i \phi}& 0\end{pmatrix}\left(\frac{1}{r}\frac{\partial}{\partial \phi}+i e\frac{B r}{2}\right)\nonumber\text{ ,}
\end{align}
only the state 

\begin{align}
\Psi_2^{\rm inner} &= \frac{1}{\mathcal{N}'''}\,e^{-1/v_F\int_R^r dr'\,(|\Delta(k_z^0, r')|+eBr'/2)}\begin{pmatrix}0\\i\\1\\0\end{pmatrix}
\end{align}
is normalizable. The state that would be connected to $\Psi_1$ is given by
\begin{align}
 \psi_1^{\rm inner} \sim e^{-1/v_{F}\int_R^r dr'\,(|\Delta(k_z^0, r')|+1/r'-eBr'/2)} \begin{pmatrix}e^{-i\phi}\\0\\0\\i\,e^{i\phi}\end{pmatrix}\text{ .}
\end{align}
This state however diverges at the origin as $\psi_1^{\rm inner} \stackrel{r\to0}{\sim} 1/r$ and is thus \textit{not normalizable}. Consequently, there is only a single normalizable zero energy state bound to the vortex. Up to the normalization, it is given by

\begin{equation}
\Psi = \Psi_2^{\rm inner} \,\Theta(R-r) + \Psi_2^{\rm outer}\,\Theta(r-R)\text{ .}
\end{equation}
When we consider a momentum $k_z$ close to $k_z^0$ or change the Zeeman gap $m$ a little bit, the system will stay in the same extended topological phase. There will thus always be a single zero energy bound state per topological momentum as long as there is no topological phase transition. We find that the Hamiltonian \eqref{eq:sc_vortex_ham} always exhibits two linearly independent zero energy bound states for $r>R$. For $M^{+\Delta}_-(k_z, r') > 0$ and $M^{-\Delta}_-(k_z, r') < 0$ at large $r$, they are given by

\begin{align}
\Psi_{+\Delta}^{\rm outer} &\sim\frac{1}{\sqrt{r}}\,e^{-\int_R^r dr'\,M^{+\Delta}_-(k_z, r')/v_F}\begin{pmatrix}e^{-i \phi}\\i\\1\\i\,e^{i\phi}\end{pmatrix}\\ 
\Psi_{-\Delta}^{\rm outer} &\sim\frac{1}{\sqrt{r}}\,e^{+\int_R^r dr'\,M^{-\Delta}_-(k_z, r')/v_F}\begin{pmatrix}e^{-i \phi}\\-i\\-1\\i\,e^{i\phi}\end{pmatrix} \text{ .}
\end{align}
The bound state will be a superposition of these two states that connects to the normalizable solution for $r<R$. The special case considered previously corresponds to $M^{+\Delta}_- = - M^{-\Delta}_- = |\Delta(r)|$ and $\Psi_2 \sim \Psi_{+\Delta} - \Psi_{\Delta-}$.


\end{document}